\documentclass[final]{article}

    \PassOptionsToPackage{numbers, compress}{natbib}
\usepackage{neurips_2024}

\usepackage[utf8]{inputenc} % allow utf-8 input
\usepackage[T1]{fontenc}    % use 8-bit T1 fonts
\usepackage{hyperref}       % hyperlinks
\usepackage{url}            % simple URL typesetting
\usepackage{booktabs}       % professional-quality tables
\usepackage{amsfonts}       % blackboard math symbols
\usepackage{nicefrac}       % compact symbols for 1/2, etc.
\usepackage{microtype}      % microtypography
\usepackage{xcolor}         % colors
\usepackage{graphicx}
\usepackage{wrapfig}
\usepackage{amsmath}
\usepackage{ulem}
\usepackage{multirow}
\usepackage{caption}
\usepackage{soul}

\title{Language Grounded Multi-agent Reinforcement Learning with Human-interpretable Communication}

\author{
  Huao Li\thanks{Work done while interning at Honda Research Institute USA. Code available at \url{https://romanlee6.github.io/langground_web/}.} \\
University of Pittsburgh \\
  \texttt{hul52@pitt.edu} \\
  \And
  Hossein Nourkhiz Mahjoub \\
  Honda Research Institute USA, Inc. \\
  \texttt{hossein\_nourkhizmahjoub@honda-ri.com} \\
  \AND
  Behdad Chalaki \\
  Honda Research Institute USA, Inc. \\
  \texttt{behdad\_chalaki@honda-ri.com} \\
  \And
  Vaishnav Tadiparthi \\
  Honda Research Institute USA, Inc. \\
  \texttt{vaishnav\_tadiparthi@honda-ri.com} \\
  \And
  Kwonjoon Lee \\
  Honda Research Institute USA, Inc. \\
  \texttt{kwonjoon\_lee@honda-ri.com} \\
    \And
  Ehsan Moradi-Pari \\
  Honda Research Institute USA, Inc. \\
  \texttt{emoradipari@honda-ri.com} \\
    \And
  Michael Lewis \\
University of Pittsburgh \\
  \texttt{ml@sis.pitt.edu} \\
    \And
  Katia Sycara \\
Carnegie Mellon University\\
\texttt{sycara@andrew.cmu.edu} 
}

\begin{document}

\maketitle

\begin{abstract}

Multi-Agent Reinforcement Learning (MARL) methods have shown promise in enabling agents to learn a shared communication protocol from scratch and accomplish challenging team tasks. However, the learned language is usually not interpretable to humans or other agents not co-trained together, limiting its applicability in ad-hoc teamwork scenarios. In this work, we propose a novel computational pipeline that aligns the communication space between MARL agents with an embedding space of human natural language by grounding agent communications on synthetic data generated by embodied Large Language Models (LLMs) in interactive teamwork scenarios. Our results demonstrate that introducing language grounding not only maintains task performance but also accelerates the emergence of communication. Furthermore, the learned communication protocols exhibit zero-shot generalization capabilities in ad-hoc teamwork scenarios with unseen teammates and novel task states. This work presents a significant step toward enabling effective communication and collaboration between artificial agents and humans in real-world teamwork settings.

\end{abstract}

\section{Introduction}

Effective communication is crucial for multiple agents to collaborate and solve team tasks. Especially in ad-hoc teamwork scenarios where agents do not coordinate a priori, the ability to share information via communication is the keystone for successful team coordination and good team performance~\cite{mirsky2020penny}. Multi-Agent Reinforcement Learning (MARL) methods have shown promise in allowing agents to learn a shared communication protocol by maximizing the task reward~\cite{sukhbaatar2016learning,singh2018learning}. However, merely maximizing the utility of communication in specific task goals might compromise task-agnostic objectives such as optimizing language complexity and informativeness~\cite{tucker2022trading,kartenintent}, making the learned communication protocols 1) hard to interpret for humans or other agents that are not co-trained together~\cite{kottur2017natural,lazaridou2018emergence,chaabouni2021communicating} and 2) highly data-inefficient~\cite{eccles2019biases}. In addition, most previous works learn communication protocols with atomic symbols or a combination of them leaving the relation between symbols unexplored~\cite{zhu2024survey}. Only a few researches attempt to learn a semantic space for zero-shot communication of unseen states~\cite{tucker2021emergent, karten2023interpretable}. One of the popular directions for interpretable communication is to regulate the learning process with external knowledge from human languages~\cite{tucker2022generalization,lazaridou2020multi,lowe2019interaction,agarwal2019community}. However, this process is challenging due to the divergent characteristics between human and machine languages by nature~\cite{chaabouni2019anti}. 
For example, most agent training methods (e.g., deep reinforcement learning) require a huge amount of data that is impractical for human-in-the-loop training or even collecting from humans~\cite{lowe2019interaction}.

The rise of Large Language Models (LLMs) provides new opportunities in grounding agent communication with human languages. Recent generative models fine-tuned with human instructions (e.g., GPT-4, Llama 3) show reasonable capabilities in completing team tasks and communicating in a human-like fashion via embodied interaction~\cite{park2023generative,li2023theory}. Essentially, LLMs encapsulate a highly trained model of human language patterns in teamwork, allowing them to generate descriptions and responses that are well-grounded in natural language. For the purpose of guiding multi-agent communication, they represent the most generally available reference based on a vast corpus of human language data that would be infeasible to collect through other means. However, LLMs are known to suffer from a lack of grounding with the task environments (a.k.a. hallucinations), which prevents embodied agents from generating actionable plans~\cite{mahowald2023dissociating}. While attempts have been made to ground LLMs with reinforcement learning or interactive data collected from environments~\cite{xiang2024language,carta2023grounding,tan2024true}, none of them involve teamwork nor communication among multiple embodied agents.

In this paper, we propose LangGround, a novel computational pipeline for artificial agents to learn human-interpretable communication for ad-hoc human-agent teamwork.
Specifically, we use synthetic data generated by LLM agents in interactive teamwork scenarios to align the communication space between MARL agents with human natural language. Learning signals from both language grounding and environment reinforcement regulate the emergence of a communication protocol to optimize both team performance and alignment with human language. We have also evaluated the learned communication protocol in ad-hoc teamwork scenarios with unseen teammates and novel task states. The aligned communication space enables translation between high-dimensional embeddings and natural language sentences, which facilitates ad-hoc teamwork. The proposed computational pipeline does not depend on specific MARL architecture or LLMs and should be generally compatible. We have sought to minimize the influence of prompt engineering to ensure the seamless applicability of our approach in diverse environments. To the best of our knowledge, this work is among the very first attempts at training MARL agents with human-interpretable natural language communication and evaluating them in ad-hoc teamwork experiments.
% 

% Firstly, the communication representations are different, with artificial agents using continuous vectors to share information while natural human languages are based on discrete tokens (e.g., words). Secondly, humans and agents must establish a common understanding of the symbols being communicated in order to coordinate successfully. Lastly, 

% Modern LLMs have the potential to produce high-fidelity synthetic human data for training agents with interpretable communication. 

% Despite the debate on whether these capabilities root from internal world models or surface statistics, m

\section{Related Work}

\subsection{Multi-Agent Communication}

% Deep learning methods have been used to simulate the emergence of language among artificial agents. 
% Multiple agents are put into interactive game settings and tasked to optimize their communication protocol in order to achieve shared goals and maximize team performance.  Essential assumptions of emergent communication are 1) agents can only exchange information via explicit communication channels (e.g. discrete symbols) but may not access other agents’ weights or gradients, 2) the communication protocol is learnt over interactions instead of pre-defined with expert knowledge. 

%researchers propose fully decentralized methods by grounding communication on input representations. 
%Results show that separately learned communication protocols without considering environment rewards perform better than previous end-to-end methods. 

Reinforcement learning has been used to coordinate the teamwork and communication among multiple agents in partially observable environments. In earlier works such as DIAL~\cite{foerster2016learning}, CommNet~\cite{sukhbaatar2016learning}, and IC3Net~\cite{singh2018learning}, agents learn differentiable communication in an end-to-end fashion under the pressure of maximizing task reward. Other works use shared parameters~\cite{karten2023interpretable} or a centralized controller~\cite{niu2021multi} to stabilize the non-stationary learning process of multi-agent communication. More recently, representation learning methods such as autoencoder~\cite{lin2021learning} and contrastive learning~\cite{lo2023learning} are used to ground an agent's communication on individual observations. However, comm-MARL methods usually suffer from overfitting to specific interlocutors trained together~\cite{lazaridou2020emergent}. The learned communication protocols can not be understood by unseen teammates in ad-hoc teams, let alone another human.

Another relevant trend of research is Emergent Communication (EC), where researchers focus more on simulating the development of natural (i.e., symbolic) language with artificial agents~\cite{lazaridou2020emergent,karten2023role}. The most common task scenarios used in the EC community are reference games or Lewis signaling games~\cite{lewis2008convention}, in which a speaker must describe an object to a listener, who must then recognize it among a set of distractions~\cite{tucker2021emergent}. However, previous research has shown that learning EC in more complicated, scaled-up, and multi-round interactive task scenarios can be challenging or even infeasible~\cite{eccles2019biases, chaabouni2021emergent}. Even in situations where agents can learn to communicate, the learned protocols are usually either not human-interpretable~\cite{kottur2017natural} or semantically drifting from human language~\cite{lee2019countering}.

% comm-MARL has less assumptions on the format of communication compared to EC. 
% By using techniques such as differentiable communication~\cite{sukhbaatar2016learning}, shared parameters~\cite{foerster2016learning,singh2018learning}, and centralized controller~\cite{niu2021multi}, comm-MARL agents are able to communicate and coordinate in challenging task scenarios including navigation~\cite{lowe2017multi}, gridworld~\cite{singh2018learning}, and StarCraft 2~\cite{zhang2019efficient}. 

\subsection{Human-Interpretable Communication}

% \huao{TO DO can extend this section a little bit: 1) explain each work in more details, 2) there are previous work pointing out the opportunity of shaping comm-MARL behaviors with a small supervised dataset~\cite{lazaridou2016multi,tucker2022generalization}. 3) will need to talk about langauge shifting}

To address the above-mentioned challenges, several recent works propose human-interpretable communication in RL settings. Lazaridou et al.\cite{lazaridou2020multi} leverage pre-trained task-specific language models to provide high-level guidance for natural language communication. A few other works align low-level communication tokens with human language\cite{lazaridou2016multi,tucker2022generalization}, or learn discrete prototype communication in a semantically meaningful space~\cite{tucker2021emergent,karten2023interpretable}.
But as pointed out in several studies~\cite{dessi2022communication,yao2022linking}, the low mutual intelligibility between human language and neural agent communication makes the alignment process non-trivial.
% evaluate the capabilities of neural models when they are provided with human-generated inputs versus language-resembling messages generated by other neural models in tasks such as caption-based image retrieval. These studies find it misleading to rely on the resemblance between these two types of inputs. 
%Some other works are evaluating the capabilities of neural models when fed by human-generated inputs versus when utilizing  natural language-resembling messages generated by other neural models, in tasks such as caption-based image retrieval, and find it deceptive to rely on this resemblance\cite{dessi2022communication}. 
Our work is closest to \cite{lowe2019interaction}, in which researchers alternate imitating human data via supervised learning and self-play to maximize reward in a reference game. The differences are that in \cite{lowe2019interaction} authors try to train neural agents for reference games in an end-to-end fashion with backpropagation, while we train MARL agents in interactive team tasks. The exploration of this research direction is still very limited, as no previous work has ever evaluated natural language communication agents within interactive task environments and for ad-hoc human-agent teams.
% 
% \huao{a few more papers to include:}
% \cite{dessi2022communication,dessi2021interpretable,,li2022learning}
\subsection{Language-Grounded Reinforcement Learning}

Reinforcement learning is known to struggle with long-horizon problems with sparse reward signals~\cite{mnih2015human}. Natural language guidance has been used to provide auxiliary rewards to improve the data efficiency and learning robustness~\cite{waytowich2019grounding}. Goyal et al.\cite{goyal2019using} use step-by-step natural language instructions provided by human annotators to construct auxiliary reward-learning modules, encouraging agents to learn from expert trajectories. Narasimhan et al.\cite{narasimhan2018grounding} research the impact of language grounding on representation learning and transfer learning of RL agents in a 2D game environment. Additional work has explored grounding RL with other formats of materials such as game manuals~\cite{hanjie2021grounding, wu2024read} and human commands~\cite{xu2022grounded}. However, none of those works has ever used the communication messages as the language ground nor guided the information-sharing process in multi-agent teamwork.

The most relevant research to our proposed method is CICERO, which empirically evaluates the proposed AI agent in the complex natural language strategy game Diplomacy with human players~\cite{meta2022human}. CICERO has an RL module for strategic reasoning and a language model for generating messages. The two modules are trained separately on different datasets, namely self-play trajectories and conversation data collected from human players. The two modules in CICERO function independently, with the only connection being that the language model takes intention estimation from the planning module as input. While in our work, both action and communication are generated by individual RL agents that are trained end-to-end with a combination of RL loss and supervised learning loss. 

% \huao{TO DO missing lit: Language Grounding for Robotics}

\section{Preliminaries}

We formulate the problem as a decentralized partially observable Markov Decision Process~\cite{oliehoek2016concise}, which can be formally defined by the tuple $(\mathcal{I},\mathcal{S},\mathcal{A},\mathcal{C},\mathcal{T}, \Omega,\mathcal{O},\mathcal{R},\gamma)$, where $\mathcal{I}$ is the finite set of $n$ agents, $s\in\mathcal{S}$ is the global state space, $\mathcal{A} = \times_{i\in\mathcal{I}} \mathcal{A}_i $ is the set of actions, and $\mathcal{C}=\times_{i\in\mathcal{I}}\mathcal{C}_i $ is the set of communication messages for each of $n$ agents. 
$\mathcal{T}: \mathcal{S} \times \mathcal{A} \to \mathcal{S}$ is the transition function that maps the current state $s_{t}$ into next state $s_{t+1}$ given the joint agent action. 
In our partially observable environments, each agent receives a local observation $o^i\in\Omega$ according to observation function $\mathcal{O}:\mathcal{S}\times\mathcal{C}\times\mathcal{I}\to\Omega$. 
Finally, $\mathcal{R}:\mathcal{S} \times \mathcal{A}\to \mathbb{R}$ is the reward function, while $\gamma\in[0,1)$ is the discount factor. 
At each time stamp $t$, each agent $i$ takes an action $a_{t}^{i}$ and sends out a communication message $c_{t}^{i}$ after receiving the partial observation of task state $s_t$ along with all messages sent by other agents from last time stamp $c_{t-1}$. 
% \VT{Maybe we can include the action function here. $\pi_i (o^i, c^i)$ or something in that format}
Each agent then receives the individual reward $r_t^i \in \mathcal{R}(s_t, a_t)$. We consider fully cooperative settings in which the reinforcement learning objective is to maximize the total expected return of all agents:
% In this work, we parameterize the policy function using a deep neural network, and use REINFORCE \cite{williams1992simple} to optimize it.

%We formulate the problem as a decentralized partially observable Markov Decision Process~\cite{littman1994markov}, which can be formally defined by the tuple $(\mathcal{S},\mathcal{A},\mathcal{C},\mathcal{T},\mathcal{O},\mathcal{R},\gamma)$. $\mathcal{S}$ is the set of states, $\mathcal{A}_i$ is the set of actions, and $\mathcal{C}_i$ is the set of communications for each of $N$ agents.  $\mathcal{T}$ is the transition function that maps the current state $s_{t}$ and next state $s_{t+1}$ given the joint agent action $\mathcal{T}: \mathcal{S} \times \mathcal{A}_1,...,\mathcal{A}_N \to \mathcal{S}$. In our partial observable environments, $\mathcal{O}$ is the set of observations which are functions of state $\mathcal{S}$ and communication $\mathcal{C}$. Finally, $\mathcal{R}$ is the reward function and $\gamma$ is the discount factor. In this work, we parameterize the policy function using a deep neural network, and use REINFORCE \cite{williams1992simple} to optimize it.

% \VT{check equation again. Reward model is a function of state, not observation - correct? Also, ${\pi^i: \mathcal{S} \to \mathcal{A} \times \mathcal{C}}$ implies that the comm policy is the same as action policy. That does not seem correct. Action policy maps observation and comms to action, whereas comm policy just maps observation to comm message.} 
\begin{equation}
    \max\limits_{\pi^i: \Omega \to \mathcal{A} \times \mathcal{C}} \mathbb{E} \left[ \sum_{t \in T} \sum_{i \in \mathcal{I}} \gamma^t \mathcal{R}({{s}}^i_t, a^i_t) | a^i_t \sim \pi^i, o^i_t \sim \Omega \right]
\end{equation}

% \begin{equation}
%     \max\limits_{\pi: \mathcal{S} \to \mathcal{A} \times \mathcal{C}} \mathbb{E} [ \sum_{t \in \mathcal{T}} \sum_{i \in N} \gamma \mathcal{R}(s_t, a_t) | (s_t, a_t) \sim \pi^i, s_t \sim \mathcal{T}(s_{t-1})]
% \end{equation}

We borrow the definition of language learning from \cite{lowe2019interaction}. We define a target language $L* \in \mathcal{L}$ that we want the agents to learn, assuming $\mathcal{L}$ is the set of natural language and $L*$ is the optimal communication language for achieving a good team performance in the specific task. Specifically, we consider a language $L \in \mathcal{L}$ to be a set of communication messages $\mathcal{C}$ which are mapped from agent observations to communication messages defined as $L: \Omega\to \mathcal{C}$. %be a set of communication messages $\mathcal{C}$ and a mapping from between agent observations and messages in the environment, $L: \mathcal{O} \times \mathcal{C}_1,...,\mathcal{C}_N \to \mathcal{C}$. 
In typical RL settings, this can be thought of as the mapping between input observation vectors and English descriptions of the observation. We consider a dataset $\mathcal{D}$ consisting of $\mathcal{|D|}$ (observation, action) pairs, which comes from expert trajectories generated by LLM embodied agents using the target language $L*$. The language learning objective is to train agents to speak language $L*$ in order to collaborate with experts in ad-hoc teamwork. It is worth noting that we want the learned language to generalize to unseen examples that are not contained in $\mathcal{D}$.

To train agents that perform well on the team task and speak human-interpretable language, we need to solve a multi-objective learning problem by combining the learning signals from both environment reward and supervised dataset $\mathcal{D}$. The process of training an optimal communication-action policy can be defined as solving an optimization problem with two constraints 1) agents must learn to communicate effectively to maximize team performance and 2) agents must learn to use similar language as in the supervised dataset $\mathcal{D}$.

% \VT{Should probably be hifen between Language Grounded (Language-grounded, not Language grounded)}
\section{Language Grounded Multi-agent Communication}

\label{pipeline}

\begin{figure}[h]
  \centering
\includegraphics[width=\textwidth,trim={0 5cm 0.5cm 0},clip]{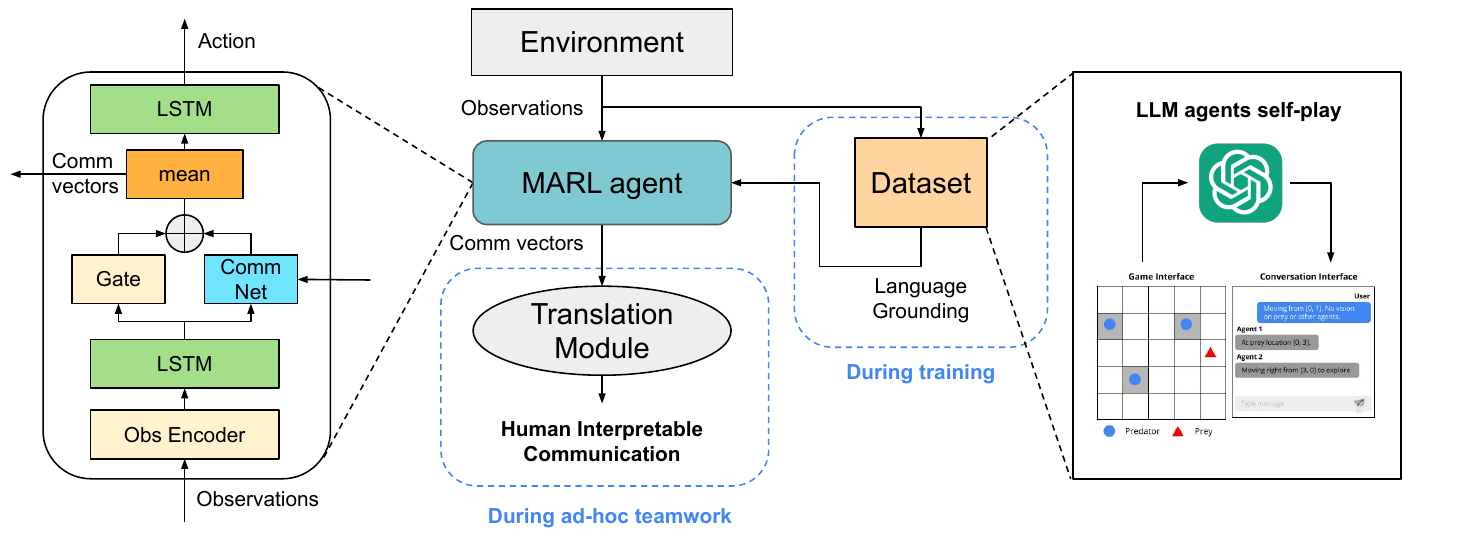}
  \caption{Illustrations of our proposed computational pipeline , LangGround. The framework consists of three modules: 1) collecting grounded communication from LLM agents, 2) aligning MARL communication with language grounds, 2) translating aligned communication vectors into natural language messages via cosine similarity matching.}
  \label{framework}
\end{figure}

In this section, we propose a computational pipeline for learning language-grounded multi-agent communication (LangGround). The general framework of LangGround is illustrated in Figure~\ref{framework} consisting of two parts: collecting grounded communication from LLM agents and aligning MARL communication with language grounds.

\subsection{Grounded communication from LLM agents}
We use embodied LLM agents to collect samples of target language $L*$. To allow LLM agents to interact with the task environments, a text interface $I$ is implemented following~\cite{li2023theory} to translate between abstract representations and English descriptions. Essentially, each of the $n$ LLM agents is provided with general prompts about the team task and instructed to collaborate with others to achieve the common goal. At each time stamp $t$, the LLM agent $i$ receives English descriptions of its own observation of the environment $I(o_t^i)$ which also includes communication messages from teammates in the last time stamp $C_{t-1}$.%and communication messages from teammates in the last time stamp $I(C_{t-1})$. 
The LLM agent is prompted to output its next action and communication message which are then encoded into abstract $\mathcal{A}_t^i$ and $\mathcal{C}_t^i$ and used to update the task environment. 

Theoretically, we construct a text environment in parallel with the actual task environment to ensure that embodied LLM agents are exposed to the equivalent information as RL agents, albeit in a different format. The action and communication policy of LLM agents emerge from the backbone LLM, since the provided prompts do not include any explicit instructions on team coordination or communication strategy. In the results section we will confirm findings from previous literature that modern LLMs (e.g., GPT-4) is able to perform reasonably well and communicate effectively in collaborative tasks. The expert trajectories generated by LLMs are used to construct the supervised dataset $\mathcal{D}$ which maps individual agent's (observation, action) pairs to natural language communication messages. $\mathcal{D}$ is used during the training of MARL to provide supervised learning signals to align the learned communication protocols with human language. More implementation details of LLM agents and data collection can be found in the Appendix.

\subsection{Multi-agent Reinforcement Learning with aligned communication}

The MARL with communication pipeline is similar to IC3Net~\cite{singh2018learning} in which each agent has an independent controller model to learn how and when to communicate. During each time-step, input observation $o_t^i$ is encoded and passed into each agent's individual LSTM. The hidden state of LSTM $h_t^i$ is then passed to the probabilistic gating function to decide whether to pass a message to other agents. A single-layer communication network transforms $h_t^i$ into communication vector $c_t^i$. The mean communication vector of all agents is finally used by each agent's LSTM to produce the action $a_t^i$.

To shape the learned communication protocol toward human language, we introduce an additional supervised learning loss during the training of MARL. Specifically, at each time step, we sample 
a reference communication from the dataset based on each agent's observation and action, $D(o_t^i,a_t^i)$, representing how a human (LLM) would communicate in the same situation. We then calculate the cosine similarity between the agent communication vector $c_t^i$ and the word embedding of the reference communication $c_h$. To align the communication space learned by MARL with the high-dimensional embedding space of natural language, we construct the supervised loss function as follows:

\begin{equation}
    L_{sup} =  \sum_{t \in T} \sum_{i \in \mathcal{I}} \left[1-cos(c_t^i, D(o_t^i,a_t^i)) \right]
\end{equation}
\begin{equation}
D(o_t^i,a_t^i) = 
\begin{cases}
    c_h & \text{if } (o_t^i,a_t^i) \in \mathcal{D} \\
    \mathbf{0} & \text{otherwise} \\
\end{cases}
\end{equation}

The construction of the communication message is shaped by two learning signals: 1) the reinforcement learning objective which determines useful information to share with other agents based on the policy loss gradient, and 2) the supervised learning objective which imitates the communication messages used by LLM agents in dataset $\mathcal{D}$. The total loss function is formulated as follows:

\begin{equation}
    L = L_{RL}+ \lambda L_{sup}
    \label{weightedLoss}
\end{equation}

The hyperparameter $\lambda$ is used to scale the supervised loss. Each agent's policy is optimized with backpropagation to minimize the joint loss.

% \begin{wrapfigure}{r}{0.5\textwidth}
%   \begin{center}
%     \includegraphics[width=0.48\textwidth,trim={0 0 12cm 0},clip]{figs/framework.pdf}
%   \end{center}
%   \caption{The construction of the communication message is shaped by two learning signals: 1) the reinforcement learning objective which determines useful information to share with other agents based on the policy loss gradient, and 2) the supervised learning objective which imitates the communication messages used by LLM agents in dataset $\mathcal{D}$. The total loss function is formulated as follows:}
%   \label{framework}
% \end{wrapfigure}

\section{Experiments}

\subsection{Environments}
In this section, we evaluate our proposed method in two multi-agent collaborative tasks with varied setups and different characteristics. The first environment, Predator Prey~\cite{singh2018learning}, is widely used in comm-MARL research as a benchmark. We include this environment to represent team tasks that require all agents to share their partial observations for the team to form a complete picture of the task state. The second environment, Urban Search \& Rescue (USAR)~\cite{li2023theory,oguntola2021deep}, presents a more demanding challenge due to the inclusion of heterogeneous team members and the temporal dependence between their behaviors. Here, agents must communicate not only their observations but also their intentions and requests to effectively coordinate. Illustrations of the two environments are shown in Figure~\ref{framework}, and more details are provided in the Appendix.

\subsection{Experiment setups}

We compare our proposed pipeline LangGround against previous methods, including IC3Net~\cite{singh2018learning}, autoencoded communication (aeComm)~\cite{lin2021learning}, Vector-Quantized Variational Information Bottleneck (VQ-VIB)~\cite{tucker2022trading}, prototype communication (protoComm)~\cite{tucker2021emergent}, and a control baseline of independent agents without communication (noComm). aeComm represents the state-of-the-art multi-agent communication methods that grounds communication by reconstructing encoded observations. It has been shown to outperform end-to-end RL methods and inductive biased methods in independent, decentralized settings. 
VQ-VIB and prototype-based method are representative solutions for human-interpretable communication, which learn a semantic space for discrete communication tokens and perform reasonably well in human-agent teams. 
Finally, IC3Net has a similar architecture to our proposed pipeline representing an ablating baseline without language grounding. 

All methods are implemented with the same centralized training decentralized execution (CTDE) architecture for a fair comparison. Each agent has an observation encoder, an LSTM for action policy, a single-layer fully-connected neural network for transferring hidden states into communication messages, and a gate function for selectively sharing messages. The parameters of the action policy and obs encoder are shared during training for a more stable learning process. We use REINFORCE~\cite{williams1992simple} to train both the gating function and policy network. The communication messages are continuous vectors of dimension $D = 256$.

\section{Results}
As for the evaluation matrices, we first consider if LangGround allows MARL agents to complete collaborative tasks successfully (i.e., task utility) and converge to a shared communication protocol quickly (i.e., data-efficiency), in comparison with other state-of-the-art methods as the baselines. Then we analyze the properties of aligned communication space such as human interpretability, topographic similarity, and zero-shot generalizability, to show how close the learned language is to the target human natural language. 
Finally, we evaluate the ad-hoc teamwork performance in which MARL agents must communicate and collaborate with unseen LLM teammates via natural language.

\subsection{Task performance}
\label{perf}

\begin{figure}

  \centering
  \includegraphics[height=0.32\textwidth]{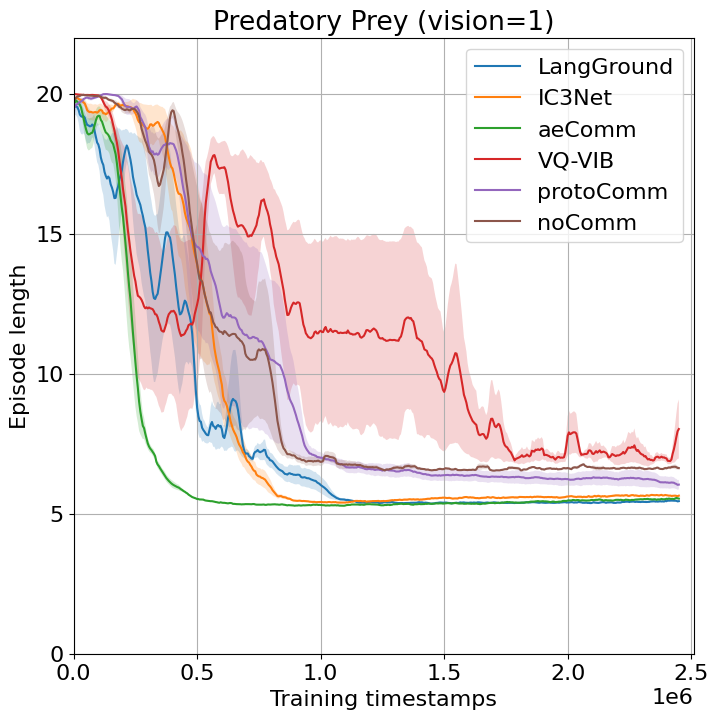}
   \includegraphics[height=0.32\textwidth]{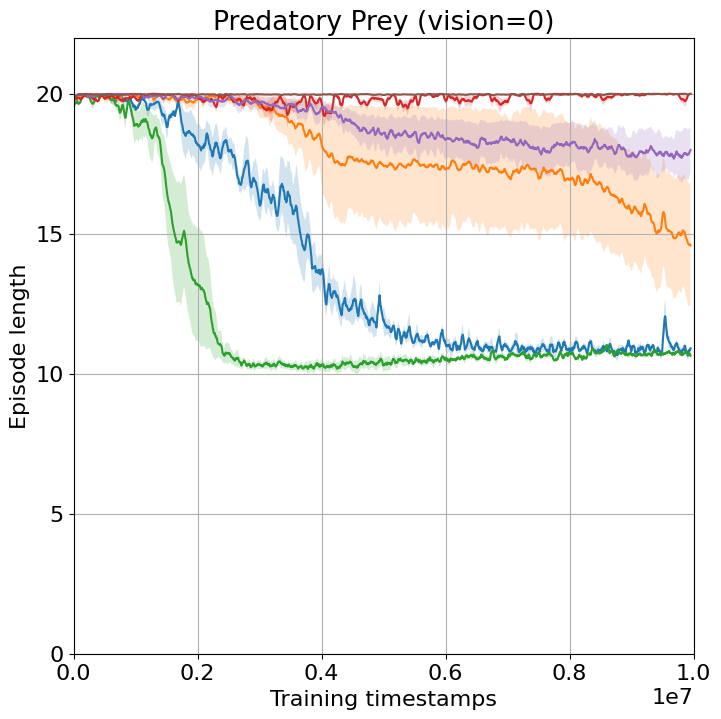}
   \includegraphics[height=0.32\textwidth]{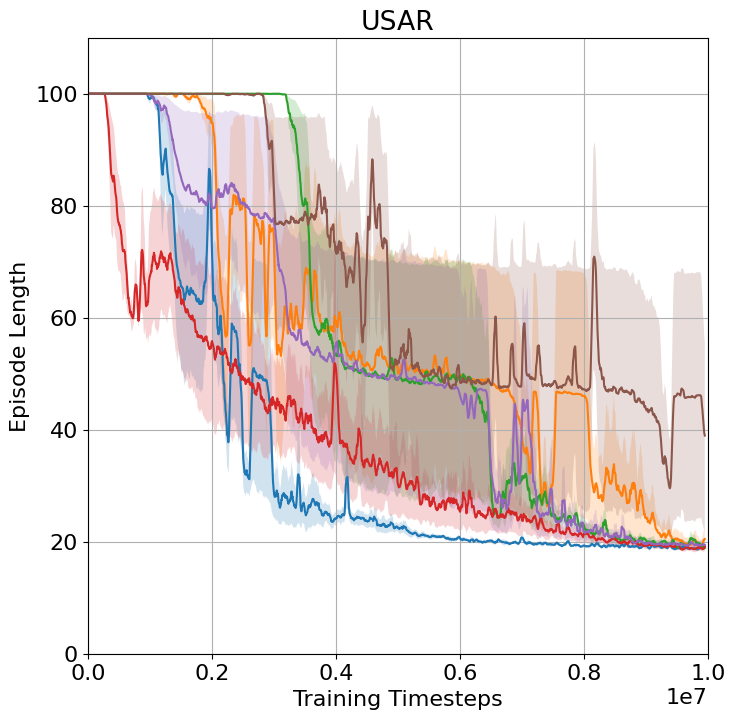}
  \caption{Learning curves of LangGround in comparison with baseline methods. The y-axis is task performance measured by the episode length until task completion, which is lower the better. The x-axis is the number of training timestamps. Shaded areas are standard errors over three random seeds.}
  \label{task_performance}
\end{figure}

In Figure~\ref{task_performance}, we compare the task performance of multi-agent teams using different communication methods by
plotting out the average episode length during training over 3 random seeds with standard errors.

In Predator Prey vision = 1 (i.e., $pp_{v1}$), our method LangGround achieves a similar final performance with IC3Net and aeComm, outperforming other baselines. However, the improvement is not outstanding due to the simplicity of the task environment. 
In Predator Prey vision = 0 (i.e., $pp_{v0}$), the predator's vision range is limited to their own location making effective information sharing more important in solving this search task. As shown in the middle figure, LangGround has a comparable final performance with aeComm and outperforms other baselines. Finally, in the most challenging USAR environment, LangGround outperforms all baselines in solving the task in fewer steps with the same amount of training time-steps. In addition, language grounding also stabilizes the communication learning process such that the variance of LangGround is much smaller than other methods. 

To test the performance of proposed method in scaled environments, we ran additional experiments in Predator Prey with a larger map size (i.e., $pp_{v1}$ (10 by 10)). The learning curves of LangGround and baselines are presented in Fig~\ref{10by10}. As shown in the figure, our method outperforms ablating baselines without language grounding (i.e., IC3Net) or without communication (i.e., noComm). This result demonstrates the benefit of introducing LangGround in stabilizing the learning process of emergent communication of MARL agents in scaled environments.

In summary, LangGround enables multi-agent teams to achieve on-par performance in comparison with SOTA multi-agent communication methods. Introducing language grounds as an auxiliary learning objective does not compromise the task utility of learned communication protocols while providing interpretability.

\begin{figure}
  \centering
  \includegraphics[width=\textwidth,trim={0 5cm 0 0},clip]{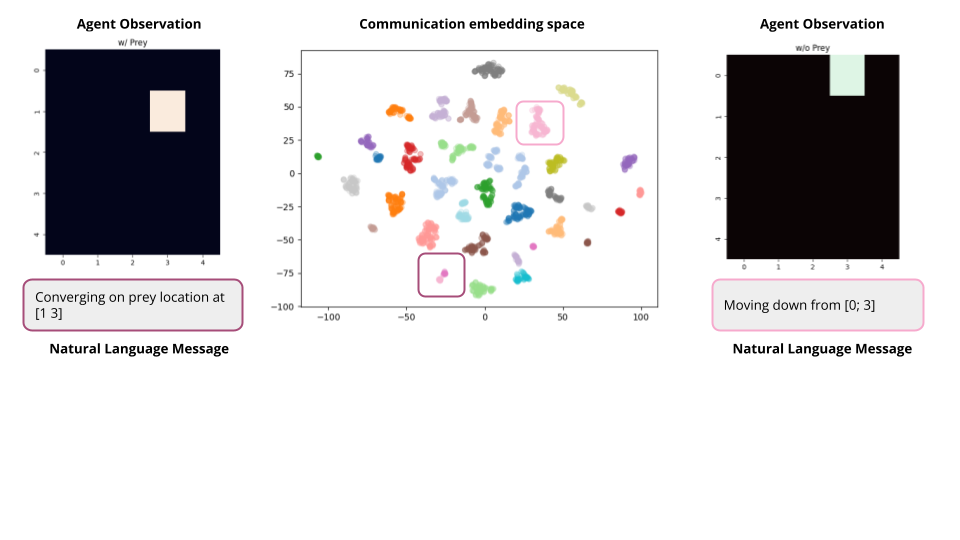}
  \caption{Learned communication embedding space. Communication vectors between agents in $pp_{v0}$ are visualized with t-SNE and clustered with DBSCAN. Two semantically meaningful clusters are identified as examples, each corresponding to a specific agent observation. We also present the most similar reference message from dataset $\mathcal{D}$ to illustrate the alignment between the agent communication space and the human language embedding space. }
  \label{space_vis}
\end{figure}

\subsection{Aligned communication space}
In addition to task utility, we are also interested in other properties of the learned communication space, such as human interpretability, topographic similarity, and zero-shot generalizability.

\subsubsection{Semantically meaningful space}

To evaluate whether the learned communication space is semantically meaningful, we visualize the learned communication embedding space by clustering message vectors sent by agents over 100 evaluation episodes in $pp_{v0}$, following~\cite{lin2021learning}. The high dimensional vectors are reduced to a two dimension space via t-SNE~\cite{van2008visualizing}, and clustered with DBSCAN~\cite{ester1996density}. As shown in Figure~\ref{space_vis}, agent communication messages can be clustered into several classifications with explicit meanings associated with agent observation from the environment. For example, the pink cluster on the right side corresponds to the situation where the agent locates in coordinates (0, 3) without vision of prey. We can look up from dataset $\mathcal{D}$ for the reference message that has the most similar word embedding with agent communication vectors in the pink cluster. The reference message (i.e., "moving down from (0, 3)") accurately refers to the agent observation, indicating the learned communication space is semantically meaningful and highly aligned with natural language embedding space.

% \begin{wrapfigure}{r}{0.4\textwidth}
%   \captionof{table}{Topographic similarity in $pp_{v0}$}
%   \label{alignMeas}
%   \centering
%   \begin{tabular}{ll}
%     \toprule
%     Methods    &  Topo Sim $\rho$\\
%     \midrule
%     LangGround  
%  & \textbf{0.67$\pm$0.07}  \\
%     IC3Net     
%  &0.54$\pm$0.14  \\
%      aeComm 
%  & 0.37$\pm$0.05  \\
%     protoComm    
%  &0.35$\pm$0.35  \\

%     \bottomrule
%   \end{tabular}
% \end{wrapfigure}

\begin{figure}[htbp]
  \centering

  \begin{minipage}[b]{0.45\textwidth}
    \centering
\begin{tabular}{llll}
\toprule
Env& $\Delta$ Cos sim & $\Delta$ Bleu score\\
\midrule
$pp_{v1}$ & 0.82$\pm$0.02 & 0.52$\pm$0.03\\
$pp_{v0}$ & 0.81$\pm$0.03 & 0.45$\pm$0.12\\
$USAR$ & 0.79$\pm$0.12 & 0.42$\pm$0.04\\
\bottomrule
\end{tabular}
    \captionof{table}{Similarity gain w/ LangGround}
    \label{alignment-table}
  \end{minipage}
\hfill
  \begin{minipage}[b]{0.45\textwidth}
    \centering
\begin{tabular}{llll}
\toprule
Grounding&  Cos sim & Bleu score\\
\midrule
100\% & 0.775 & 0.633\\
75\% & 0.667 & 0.498\\
50\% & 0.304 & 0.308\\
25\% & 0.188 & 0.224\\
\bottomrule
\end{tabular}

    \captionof{table}{Zero-shot generalization evaluation results on $pp_{v1}$ (10 by 10).}
    \label{gen_10by10}
  \end{minipage}

\end{figure}

% \begin{wrapfigure}{r}{0.6\textwidth}

% \captionof{table}{Similarity gain w/ LangGround}
% \label{alignment-table}
% \centering
% \begin{tabular}{llll}
% \toprule
% Measures& $pp_{v1}$ & $pp_{v0}$& $USAR$\\
% \midrule
% $\Delta$ Cos sim  & 0.82$\pm$0.02 & 0.81$\pm$0.03 & 0.79$\pm$0.12  \\
% $\Delta$ Bleu score & 0.52$\pm$0.03 & 0.45$\pm$0.12 & 0.42$\pm$0.04  \\
% \bottomrule
% \end{tabular}

% \end{wrapfigure}

\subsubsection{Human interpretability}
Given the goal of aligning agent communication with human language, it is intuitive to evaluate the human interpretability of language-grounded agent communication. We use the offline dataset $\mathcal{D}$ as the reference to calculate the similarity between communication messages shared by LangGround agents and LLM agents in same situations. Given 100 evaluation episodes in $pp_{v0}$, we calculate 1) the cosine similarity between word embedding and agent communication vectors, and 2) BLEU score between natural language messages and reference messages in $\mathcal{D}$ with the most similar word embedding as agent communication vectors. The results are shown in Table~\ref{alignment-table}, demonstrating that LangGround achieves significant gains in both metrics, cosine similarity and BLEU score, compared to baselines without alignment. These measures for other baseline methods would be equivalent to random chance because none of them are grounded with language during training, and hence we do not compute their performance on these metrics.

% We provide the results for IC3Net as a reference where the cos sim and bleu score are both very low.

% \begin{figure}
%   \centering
%   \includegraphics[height=0.3\textwidth]{figs/v1_alignment.png}
%    \includegraphics[height=0.3\textwidth]{figs/v0_alignment.png}
%    \includegraphics[height=0.3\textwidth]{figs/USAR_alignment.png}
%   \caption{1}
%   \label{alignment}
% \end{figure}

\subsection{Ad-hoc Teamwork}
The ultimate goal of our proposed pipeline is to facilitate ad-hoc teamwork between unseen agents without pre-coordination. Here, we propose two experiments to evaluate the zero-shot generalizability and ad-hoc collaboration capability of our trained agents.

\subsubsection{Zero-shot generalizability}
One prerequisite of ad-hoc teamwork is the ability to communicate about unseen states to their teammates. We evaluate this capability by removing a subset of prey spawn locations from environment initialization of $pp_{v0}$ and training LangGround agents from scratch. In this condition, agents would neither be exposed to nor receive any language grounding for those situations during training. During the evaluation, we record the communication messages used by agents in those unseen situations and compare them with ground truth communication from dataset $\mathcal{D}$. Results show that agents are still able to complete tasks when the prey spawns in those 4 unseen locations. As shown in Table~\ref{zeroshot-table}, the communication messages agents used to refer to those unseen locations are similar to natural language sentences generated by LLMs.

\begin{figure}[!htbp]
  \centering
  \begin{minipage}[b]{0.32\textwidth}
    \centering
        \includegraphics[width=\textwidth]{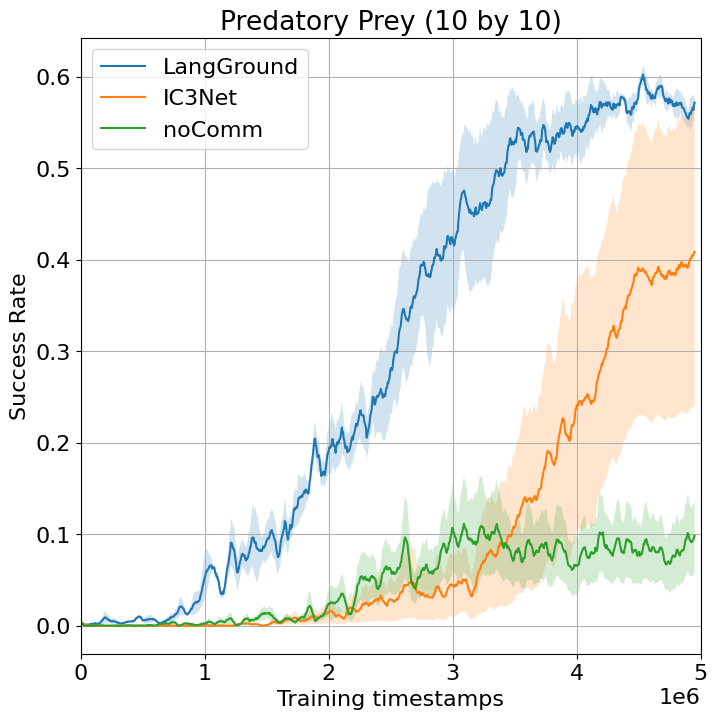}
    \captionof{figure}{Team performance of LangGround and baselines on Predator Prey with 10 by 10 map and vision range of 1.}
    \label{10by10}
  \end{minipage}
  \hfill
  \begin{minipage}[b]{0.32\textwidth}
    \centering
\includegraphics[width=\textwidth]{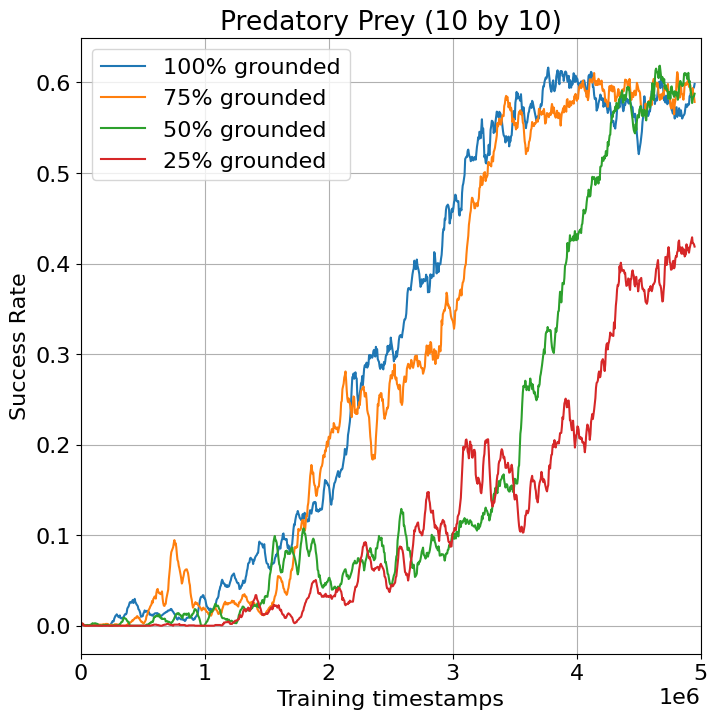}
    \captionof{figure}{Team performance of LangGround agent with different levels of language grounding on $pp_{v1}$ (10 by 10).}
    \label{10by10_perf}
  \end{minipage}
    \hfill
  \begin{minipage}[b]{0.32\textwidth}
    \centering
\includegraphics[width=\textwidth]{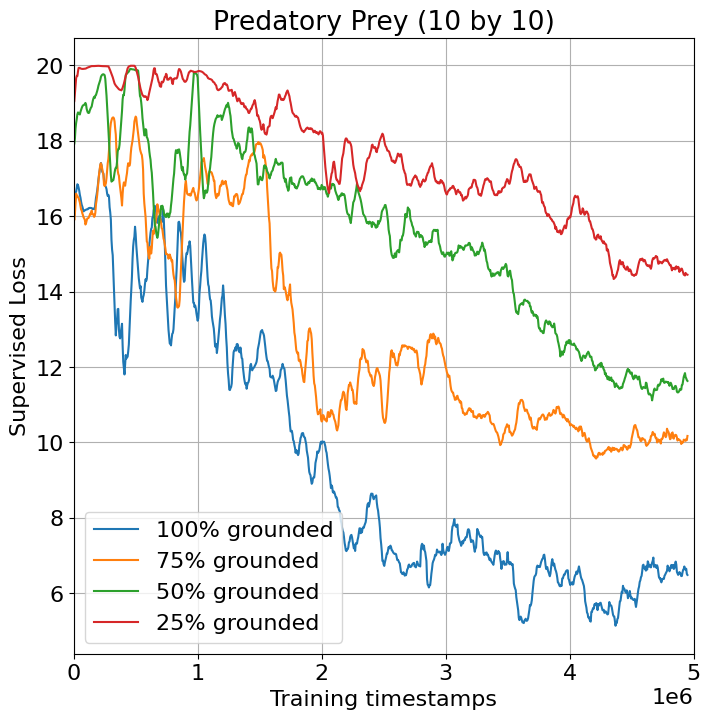}
    \captionof{figure}{Communication alignment of LangGround agent with different levels of language grounding on $pp_{v1}$ (10 by 10).}
    \label{10by10_loss}
  \end{minipage}
\end{figure}

Additionally, we train LangGround agents on Predator Prey (10 by 10) but only provide language grounding in a subset of states (i.e., 25\%, 50\%, 75\%, 100\%). The learning curves of LangGround agents with different levels of language grounding are presented in Fig.~\ref{10by10_perf} and Fig.~\ref{10by10_loss}. As shown in the figures, the more grounded states, the better the team performance, as well as the better the communication alignment. Table~\ref{gen_10by10} shows similar results of communication alignment in un-grounded states during evaluation. The first column refers to percentage of grounding data used during training. The other two columns are alignment measurements when agents encounter un-grounded states during evaluation.

To summarize, these findings confirm the alignment between the agent communication space and the human language word embedding space in zero-shot conditions. More importantly, we show that LangGround is not merely a memorization of one-to-one mapping between observations and communications but also shapes the continuous communication space in a semantically meaningful way. We could ground the agent communication space with the word embedding space of human language on a limited number of instances, and expect LangGround to output interpretable messages in un-grounded states via topographic similarity (discussed more in the Appendix) of the aligned space. In practice, this assumption depends on many factors such as the coverage of offline dataset, number of grounded states, dimension of the communication space, scale of the problem, etc. The above experiments illustrate the impact of language grounding percentage, and we leave further investigation to future work.

% ungrounded \VT{do you mean previously unseen locations} locations. 
%  distribute across the embedding space following topographic similarity, that similar messages refer to close locations. 
% {\color{red}For example, green cell is adjacent to red cell, so as their corresponding messages.} \VT{unclear sentence.}
% In addition, if we translate those messages into natural language using $\mathcal{D}$, most of them refers to locations adjacent to the actual prey locations. 

%\huao{consideing evaluating zero-shot generlizability with sending message only, instead of evaluating task performance}
% \begin{figure}
% \label{generliza}
%   \centering
%   \includegraphics[width=\textwidth]{figs/v0_generalize.png}

%   \caption{Generalization experiment results with 5 ungrouned prey locations and corresponding communication messages in the embedding space.}
% \end{figure}

\begin{table}[]
  \caption{Zero-shot generalizability in $pp_{v0}$}
  \label{zeroshot-table}
  \centering
  \begin{tabular}{llll}
    \toprule
    Prey Loc    & Cos sim     & Bleu score & Example message\\
    \midrule
    (1,1) & 0.81  & 0.41
& Moving up to converge on prey location at (1,0) for capture  \\
    (1,3)     & 0.81  & 0.27 
 & Converging on prey location at (1,3) \\
     (3,1) & 0.82  & 0.51 
 &  Moving up toward prey location at (3,1)\\
    (3,3)     & 0.78  & 0.72 
 &Converging on prey location at (3,3)  \\

    \bottomrule
  \end{tabular}
\end{table}

\subsubsection{Ad-hoc teamwork between MARL and LLM agents}
\label{adhoc}
Finally, we evaluate the performance of our agents to work with unseen teammates in ad-hoc teamwork settings. Ad-hoc teamwork refers to situations where agents collaborate with unseen teammates without pre-coordination. In this work, we use embodied LLM agents to emulate human behaviors in human-agent ad-hoc teams. Teams with 2 MARL agents and 1 LLM agent were evaluated on 8 episodes over 3 random seeds, resulting in a total of 24 evaluation episodes per condition. Ad-hoc teamwork performance is measured by the number of steps taken to complete the task; therefore, lower is better. Means and standard deviations of each condition are reported in Table~\ref{adhoc-table}.
% The ad-hoc team performance of MARL and LLM agents is shown in Table~\ref{adhoc-table}. 

\begin{table}[h]
\caption{Ad-hoc teamwork performance (lower is better)}
\label{adhoc-table}
\centering
\begin{tabular}{lllll}
\toprule
Team composition& $pp_{v1}$ & $pp_{v0}$ & $USAR$ \\
\midrule
LangGround & 4.3 $\pm$ 1.20 & 10.9 $\pm$ 4.53 & 22.0 $\pm$ 4.24 \\
LLM & 6.8 $\pm$ 5.20 & 11.6 $\pm$ 5.30 & 15.9 $\pm$ 3.37 \\
 \hline

LangGround + LLM & \textbf{8.5 $\pm$ 5.76} & \textbf{15.5 $\pm$ 4.80} & 23.2 $\pm$ 10.61 \\
aeComm + LLM & 10.3 $\pm$ 6.46 & 17.5 $\pm$ 4.60 & 20.3 $\pm$ 9.07 \\
noComm + LLM & 10.6 $\pm$ 5.73 & 20.0 $\pm$ 0.00 & 32.4 $\pm$ 13.47 \\

\bottomrule
\end{tabular}
\end{table}

We find that 1) homogeneous teams (i.e., LangGround and LLM) achieve better performance than ad-hoc teams because of their common understanding of both action and communication. As those agents are either co-trained together or duplicates of the same network, they form a stable strategy for team coordination and information sharing. Since ad-hoc teams (e.g., LangGround + LLM) were not trained together nor speak the same language, their decreased performance is expected. 2) The ad-hoc team performance of LangGround agents is better than noComm and aeComm agents in at least two out of three evaluation scenarios. Because aeComm is not aligned with human language, it serves as a baseline with a coordinated action policy and a random communication policy. The advantage of our method over aeComm and noComm merely comes from effective information sharing via the aligned language with unseen teammates. The empirical evidence presented in this section confirms the application of our method in ad-hoc teamwork.

\section{Discussion}
In this work, we developed a novel computational pipeline to enhance the capabilities of MARL agents to interact with unseen teammates in ad-hoc teamwork scenarios. Our approach aligns the communication space of MARL agents with an embedding space of human natural language by grounding agent communications on synthetic data generated by embodied LLMs in interactive teamwork scenarios.

Through extensive evaluations of the learned communication protocols, we observed a trade-off between utility and informativeness. According to the Information Bottleneck principle~\cite{tucker2022trading}, informativeness corresponds to how well a language can be understood in task-agnostic situations, while utility corresponds to the degree to which a language is optimized for solving a specific task. By introducing the additional supervised learning signal, our method pushes the trade-off toward informativeness compared to traditional comm-MARL methods that merely optimize for utility. This partially explains the "inconsistent" patterns we observe in Figure~\ref{task_performance} and Table~\ref{adhoc-table} across different task scenarios. In relatively easy tasks such as predator-prey, the learned communication is more optimized for informativeness, aligning better with human language and generalizing better in ad-hoc teamwork. In more challenging tasks such as USAR, the learned communication is more shaped toward task utility, resulting in faster convergence but less interpretability to unseen teammates.

Additionally, we found that introducing language grounding does not compromise task performance but even accelerates the emergence of communication, unlike the results reported in previous literature, where jointly optimizing communication reconstruction loss and RL loss leads to a drop in task performance~\cite{lin2021learning,lo2023learning}. The main reason for this contradiction is that our dataset $\mathcal{D}$ consists of expert trajectories from LLM embodied agents with a well-established grounding on the task. Therefore, the language grounding loss not only shapes the communication but also guides the action policy of MARL agents by rewarding behavior cloning and providing semantic representations of input observations.

It worth noting that LangGround is an extremely flexible pipeline with most of the modules being interchangeable, such as the word embedding model, base MARL-comm model, message dataset source, and the translation module. This allows us to allocate appropriate tasks to LLMs and RL respectively, considering LLMs are known to have good linguistic capabilities (e.g., describing) while struggle with formal reasoning (e.g., planning). While we use embodied LLM agents to collect grounding communication datasets for LangGround as explained in Section~\ref{pipeline}, These datasets can also come from rule-based agents or human participants, as long as they show effective communication in solving collaborative tasks. Because only communication messages are used, LLM agents' hallucinations and  does not impact MARL's task performance directly. In more complicated scenarios in which communication is beyond describing observation and action, we could still expect LLMs to generate reasonable outputs. Compared to alternative methods in embodied agents where LLMs must make correct action planning at every timestamp, it is more feasible to collect semantically meaningful messages from either LLMs or any other sources.

We believe our work would benefit the broader society for the following reasons. This research provides empirical evidence of linguistic principles during language evolution among neural agents, which might provide insights for broader research communities, including computational linguistics, cognitive science, and social psychology. The usage of embodied LLM agents as interactive simulacra of human team behaviors has a broad impact since it has potential applications in social science and may deepen our understanding of modern LLMs. Most importantly, our proposed pipeline takes initial steps in enabling artificial agents to communicate and collaborate with humans via natural language, shedding light on the broad research direction of Human-centered AI.

As for future directions, we plan to evaluate our proposed pipeline in more complicated task environments at scale and experiment with different selections of MARL algorithms, backbone LLMs, and word embeddings. Particularly, we plan to replace the use of a static dataset $\mathcal{D}$ by querying LLMs online during the training of MARL. This may allow us to capture complex information exchanged among team members in addition to individual observations, such as beliefs, intents, and requests.

\begin{ack}

We would like to thank Mycal Tucker, Seth Karten and the original authors of IC3Net for providing public access to their code base that were instrumental in this research. We also appreciate the helpful discussions with Lu Wen, Yikang Gui, and Vasanth Reddy that contributed to the development of this work.

\end{ack}

\bibliography{root}
\bibliographystyle{plain}

%%%%%%%%%%%%%%%%%%%%%%%%%%%%%%%%%%%%%%%%%%%%%%%%%%%%%%%%%%%%
\newpage
\appendix

\section{Implementation details}

\subsection{Embodied LLM agents}

Large language models are prompted to interact with the task environments in team tasks. We implement embodied LLM agents based on the pipeline proposed in~\cite{li2023theory}, where agents are augmented with explicit belief state and communication for better team collaboration capability. Each agent keeps a memory of his own observations from the environment and communication messages from other team members. Exact prompts can be found in the code within the supplementary materials. The design principles are that we only provide general rules about the task environments without explicitly instructing them on any coordination or communication strategy. We attempt to minimize the influence of prompt engineering to ensure the seamless applicability of our approach in diverse environments. In our language grounding data collection and ad-hoc teamwork experiments, we use OpenAI's API to call gpt-4-0125-preview as the backbone pre-trained model and set the temperature parameter to 0 to ensure consistent outputs.

Example prompts for LLM agents in the $USAR$ environment are provided below:

\begin{quote}
    Welcome to our interactive text game! In this game, you'll assume the role of a specialist on a search and rescue team. Alongside two other players, you'll navigate a five-room environment with a mission to defuse five hidden bombs.

\textbf{The Map}: Imagine a network of rooms represented by a connected graph where each node corresponds to a room, and the edges between nodes depict hallways. The rooms are numbered 0, 3, 6, 5, and 8. Room 0 is connected to all other rooms. Room 5 shares a hallway with room 6. Room 3 is linked to room 8. And room 8 is also connected with room 6. You can only travel to adjacent, directly connected rooms at each turn.

\textbf{The Challenge}: Scattered among these rooms are five bombs, each coded with different phases represented by colors. To defuse them, you'll need to use the correct wire-cutting tools in the correct sequence. There are one-phase, two-phase, and three-phase bombs, needing 1, 2, or 3 color-coded tool applications in sequence to disarm. For instance, a bomb with a red-green phase sequence requires the red tool first, then the green one. Points are awarded based on the number of tools used for defusing a bomb, with each tool use worth 10 points. Your task is to maximize the team score as soon as possible. The challenge is that the bomb locations and sequences are unknown to players at the start.

\textbf{Tools}: Each player is equipped with two color-coded wire cutters. As player Alpha, you have red and green tools, player Bravo wields green and blue, and player Charlie possesses blue and red.

\textbf{Actions}: Each round, you can opt to do one of the following: 1) Move to an adjacent room, 2) Inspect a bomb's phase sequence in your current room, or 3) Apply your wire cutters to a bomb in the current room. 

\textbf{Communications}: In addition to selecting an action to take from the above list, you can also send communication message texts to both of your teammates in each round. The message text you sent will be shared with both of your teammates in their observation in the next round. 

\textbf{Observation}: While you can only see what's in your current room and read text messages from teammates. You'll also be informed of the current round number, team score and the current location of your teammates. Your teammates have the same observability as you. They will not be able to know your action and its consequences unless you explicitly communicate.

To facilitate our interaction, reply your action selection and communication messages in this fixed format: Action selection: Your action. Message to Team: “Your Message”. To move to an adjacent room, say: 'Move to Room X'. To inspect the sequence of a bomb in your current room, say: 'Inspect Bomb'. To apply a wire cutter tool, say: 'Apply X Tool'. Remember, your replies must adhere strictly to these rules. Feel free to ask clarifying questions if needed. I'll supply the necessary information as we progress. Are you ready to take on this explosive challenge?

\end{quote}

Example interactions between LLM agents and environments are provided below:

\begin{quote}
    \textbf{Env}: You are now in Room 3 with Bomb 5. What is your next action?
    
    \textbf{Alpha}: Action selection: Move to Room 5. Message to Team: "I am moving to Room 5 t help Charlie with Bomb 3.
    
    \textbf{Env}: You can not directly move to Room 5 because it is not adjacent to your current location; Room 3. Consider taking a detour to another room first and then move to your destination.
\end{quote}

Example communication among LLM agents in $USAR$:
\begin{quote}
\textbf{Alpha}: I am inspecting Bomb 1 in Room 0. Bravo; please move to Room 3. Charlie; please move to Room 5.

\textbf{Bravo}: I am moving to Room 3 as instructed. Alpha; please defuse Bomb 1 in Room 0 using the red tool.

\textbf{Charlie}: I am moving to Room 5 to search for more bombs.
\end{quote}

\subsection{LangGround and baseline MARL agents}

The implementation of LangGround and baseline MARL methods is based on open-sourced codebase provided by the original paper authors~\cite{tucker2022trading,kartenintent,tucker2021emergent,singh2018learning,oguntola2021deep,lin2021learning}. Agents were trained with the same setup and hyperparameters after empirical searching. The batch size is 500, and the number of update iterations in an epoch is 10. Training on $pp_{v0}$ and $USAR$ takes 2000 epochs and $1e7$ timestamps, which takes about 4 hours to complete. Training on $pp_{v1}$ takes 500 epochs and $2.5e6$ timestamps, which takes about 1.5 hours to complete. We use a learning rate of 0.0001 for $USAR$ and 0.001 for $pp$. MARL agent's action policy is an LSTM with a hidden layer of size $256$. Communication vectors are exchanged one round at each timestamp. The supervised learning weight $\lambda$ is 1 in $pp$ and $10$ in $USAR$. VQ-VIB and prototype communication agents were allowed to use $58$ different discrete tokens to share information during teamwork. All experiments were conducted on a machine with a 14-core Intel(R) Core(TM) i9-12900H CPU and 64GB memory.

\section{Environment details}

\begin{figure}[h]
  \centering
   \includegraphics[height=0.35\textwidth,trim={0 9cm 16cm 0},clip]{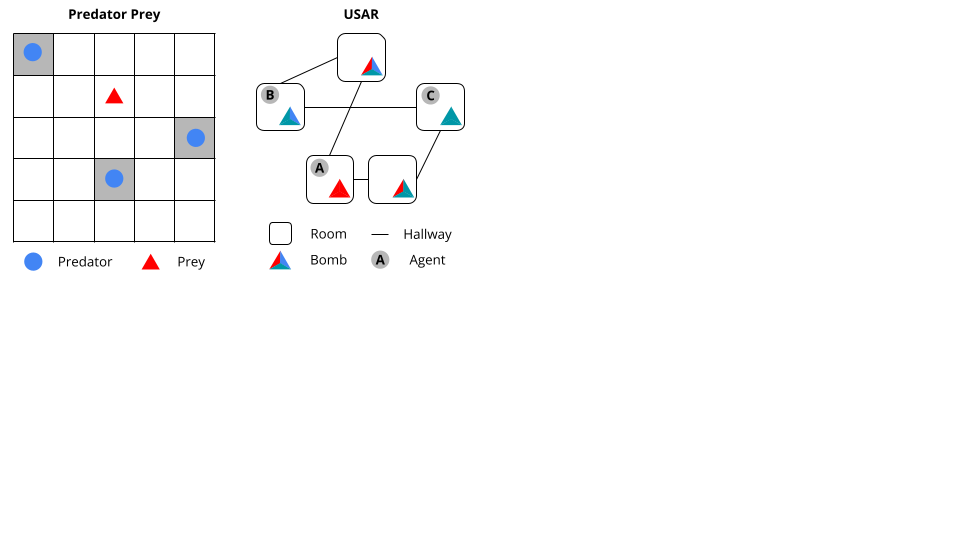}
  \caption{Illustrations of the evaluation environments. Predator Prey is a gridworld environment conceptualizing a team of predators with partial observation trying to search for a static prey. The task goal is for all predators to reach the prey location within the time limit. USAR simulates a team of specialists searching for and defusing bombs in an unknown environment. Because each specialist has the unique capability of defusing bombs in different colors, the team must coordinate to complete the task efficiently.}

\end{figure}

\subsection{Predator Prey}
\label{pp}
In this task, $n$ predators with a limited range of vision $v$ need to search for stationary prey on an $x$ by $x$ grid-world environment. Each predator receives a positive reward upon reaching the prey location. Each episode ends when all predators reach the prey or exceed the maximum number of steps $T$. The initial locations of predators and the prey might spawn anywhere on the map. At each timestamp, the predator agent receives a partial observation of $v$ by $v$ grids around its own location and may select a movement action to navigate through the map. Since this is a collaborative task, predators must learn to communicate their partial observations to allow for optimal navigation of the team. We consider Predator Prey to be a more challenging task since it has a higher-dimensional observation space and a more complex action space~\cite{lin2021learning}. Through this environment, we aim to demonstrate that aligning agents' communication with human language is a straightforward yet effective method of grounding their communication with task observations.

\subsection{USAR}
\label{usar}
The USAR task environment is designed to simulate the collaborative and problem-solving dynamics of a search and rescue mission. Three agents (i.e., Alpha, Bravo, and Charlie) need to collaborate in locating and defusing color-coded bombs hidden in an unexplored environment. Each bomb has unique phase sequences in $m$ colors, which are not revealed until inspected by agents. Agents start with different colored cutters and must use them in the correct sequence to defuse bombs. The environment is represented as a graph, where each of the $n$ rooms is a node, and the hallways connecting them are edges. At each timestamp, every agent can choose from three different types of actions: moving to one of the $n$ rooms, inspecting a bomb's sequence in the current room, or utilizing one of the $m$ wire-cutters. The size of the action space depends on the problem scale (i.e., $n+m+1$). Agents' observations are limited to their current room's contents and agent status. The team is rewarded 10*$x$ points when an $x$-phase bomb is successfully defused. An episode ends when the team has defused all bombs or exceeded the time limit. This task is designed to force team coordination since each team member has unique observations and capabilities. For example, each agent only has a subset of wire cutters and must coordinate with other teammates to defuse bombs with multiple phases. Therefore, an effective communication protocol is required for efficient information sharing and team synchronization.

\subsection{Environment configurations}
For the predator and prey environment, we use a map size of $5$ by $5$ with $3$ predators and $1$ prey. The predator's range of vision is manipulated to be either $0$ or $1$ to create two variants of the task environment. In the situation of vision = $0$, predators cannot observe the prey until they jump onto the same location. We set the maximum episode length to $20$ based on previous research~\cite{karten2023interpretable}. The USAR environment comprises five rooms ($n = 5$) and five bombs, including two single-phase, two double-phase, and one triple-phase bombs. The bomb phase might have three different colors ($m = 3$). Each of the $3$ agents spawns with $2$ different wire cutters, forcing the team to collaborate in defusing bombs with multiple phases. Each successfully defused bomb awards the team $10$ points per processed phase, resulting in $90$ as the maximum score per mission. We set the maximum episode length to $100$ based on previous research~\cite{li2023theory}.

\subsection{Text game interface}
The initial task environments of Predator Prey and USAR are implemented for MARL agents based on Gym API~\cite{1606.01540}. To facilitate interaction between LLM-based agents with the environment, we implement a rule-based text interface for each task. At each timestamp, LLM agents sequentially interact with the environment, receiving observations and performing actions via natural language interaction. Additionally, they are allowed to broadcast communication messages in natural language which are appended with the observation text and sent to all team members in the next round. It is worth noting that LLM agents receive equivalent information as MARL agents, limited to individual agent's partial observation.

The text interface facilitates communication between the game engine and the language model agents by converting game state observations into natural language descriptions and mapping agent responses back to valid game actions. To generate observations, the interface extracts relevant state features from the game engine, such as the current round number, cumulative team score, action feedback, visible objects, and communication messages from other agents. It then populates predefined sentence templates with these extracted features to produce a structured natural language description of the current game state. Action encoding relies on keyword matching, as the language models are instructed to format their responses using specific keywords and structures. The interface scans the agent's response for these predefined keywords and maps them to corresponding game actions. In cases where an agent's response is invalid or ambiguous, such as attempting to perform an action in an incorrect location, the interface generates an informative error message based on predefined rules and templates. For instance, if an agent attempts to inspect a non-existent bomb, the interface might respond with the following error message: "There is no bomb in the current location, Room X, for you to inspect." This targeted feedback helps the agents refine their actions to comply with the game's rules and constraints.

% \subsection{}

% \begin{figure}
%   \centering
%   \includegraphics[height=0.24\textwidth]{figs/lambda_loss_USAR.png}
%    \includegraphics[height=0.24\textwidth]{figs/lambda_step_USAR.png}
%    \includegraphics[height=0.24\textwidth]{figs/sup_loss_USAR.png}
%       \includegraphics[height=0.24\textwidth]{figs/steps_USAR.png}
%   \caption{1}
%   \label{alignment}
% \end{figure}

\section{Data collection details}

\subsection{LangGround dataset}
In order to construct dataset $\mathcal{D}$, we collected expert trajectories from embodied LLM agents powered by GPT-4 in interactive task scenarios. As shown in Table~\ref{adhoc-table}, teams consisting of pure LLM agents perform reasonably well in comparison to MARL methods. Therefore, we believe their action and communication policy can be used in guiding MARL agents. In $USAR$, we collected 50 episodes resulting in 2550 pairs of (observation, action) and communication messages of individual agents. The number of data pairs is 1893 for $pp_{v0}$ and 2493 for $pp_{v1}$, respectively. To facilitate the alignment of agent communication space and natural language, we use OpenAI's word embedding api (i.e., text-embedding-3-large) to translate each natural language message into a high-dimensional vector with the same dimension (i.e., $D = 256$) as agent communication vectors.
\subsection{Ad-hoc teamwork}
Due to the restrictions of resources and time, we use embodied LLM agents to emulate human behaviors in human-agent teams. We match 2 MARL agents with 1 unseen LLM agent in a team and ask them to complete the collaborative task in predator-prey and USAR environments. The LLM agent is powered by GPT-4-turbo and prompted to output action selection and communication messages given observation inputs, similar to the data collection process introduced in Section\ref{pipeline}. The Gym environment is wrapped with a text interface to decode observations into English descriptions and encode LLM agent's output into concrete action selection. Both MARL agents and LLM agents interact with the same task environment in sequence. Natural language communication messages from LLMs are embedded using OpenAI's word embedding API and sent to MARL agents. The communication vectors from MARL agents are translated to English sentences via cosine similarity matching in dataset $\mathcal{D}$.

\section{Additional Experiment Details}

\subsection{Topographic similarity}
The topographic similarity is defined by the correlation between object distances in the observation space and their associated signal distances in the communication space~\cite{brighton2006understanding}. This property is usually associated with language compositionality and ease of generalization. The intuition behind this measure is that agents should emit similar communication messages given semantically similar observations. We calculate this measure following~\cite{lazaridou2018emergence}, based on agent trajectories collected from 100 evaluation episodes in $pp_{v0}$. We first calculate 1) the cosine similarity between all pairs of communication vectors, and 2) the Euclidean distance between all pairs of agent locations. 
Then, we calculate the negative Spearman correlation $\rho$ as the measure of topographic similarity.
% Intuitively, if agents omit similar messages in nearby locations, and dissimilar messages in distant locations, the topographic similarity of their communication protocol should be high, with the highest possible value being 1. 
Table~\ref{alignMeas} indicates that our method (i.e., LangGround) results in the highest topographic similarity $\rho=0.67$ among all other baselines, exhibiting a relatively more similar property as human language.

\begin{table}[h]
    \centering
  \begin{tabular}{ll}
    \toprule
    Methods    &  Topo Sim $\rho$\\
    \midrule
    LangGround  
 & \textbf{0.67$\pm$0.07}  \\
    IC3Net     
 &0.54$\pm$0.14  \\
     aeComm 
 & 0.37$\pm$0.05  \\
    protoComm    
 &0.35$\pm$0.35  \\
    \bottomrule

  \end{tabular}
        \caption{Topographic similarity in $pp_{v0}$}
    \label{alignMeas}
\end{table}

\subsection{Additional ablation study}

Here we run additional ablation study to attribute reinforcement learning signal and supervised learning signal to the agent's action and communication output in the multi-objective optimization problem. 

Because the LangGround agent is trained end-to-end with a combination of RL and SL loss and uses the intermediate hidden state of its policy as the communication vector, it is very hard to separate the reasoning processes of action and communication. However, we could provide indirect evidence to prove that RL and SL jointly contribute to both the agent's action and communication.

\begin{figure}[!htbp]
  \centering
  \begin{minipage}[b]{0.32\textwidth}
    \centering
        \centering
        \includegraphics[width=\textwidth]{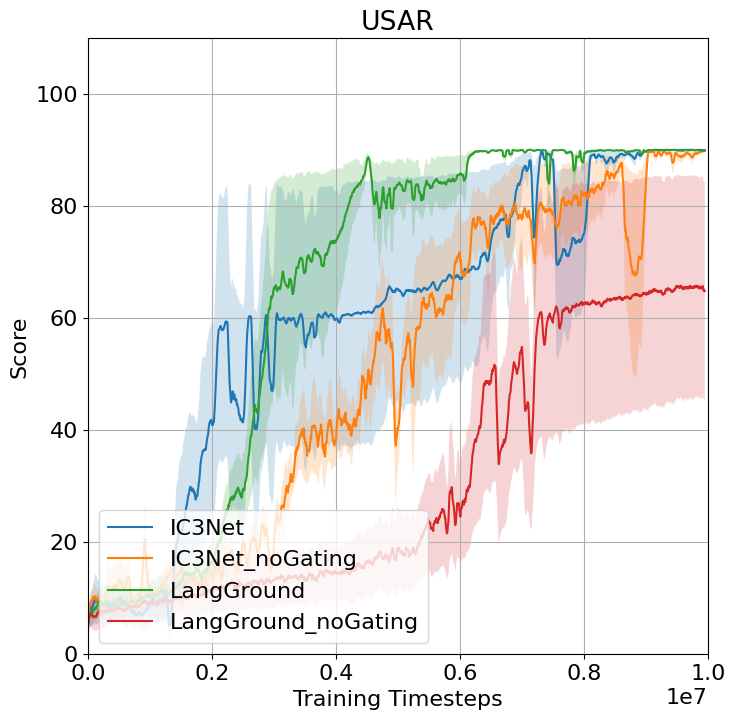}
    \captionof{figure}{Ablation study of gating function and language grounding. }
    \label{removeGating}
  \end{minipage}
  \hfill
  \begin{minipage}[b]{0.32\textwidth}
    \centering
\includegraphics[width=\textwidth]{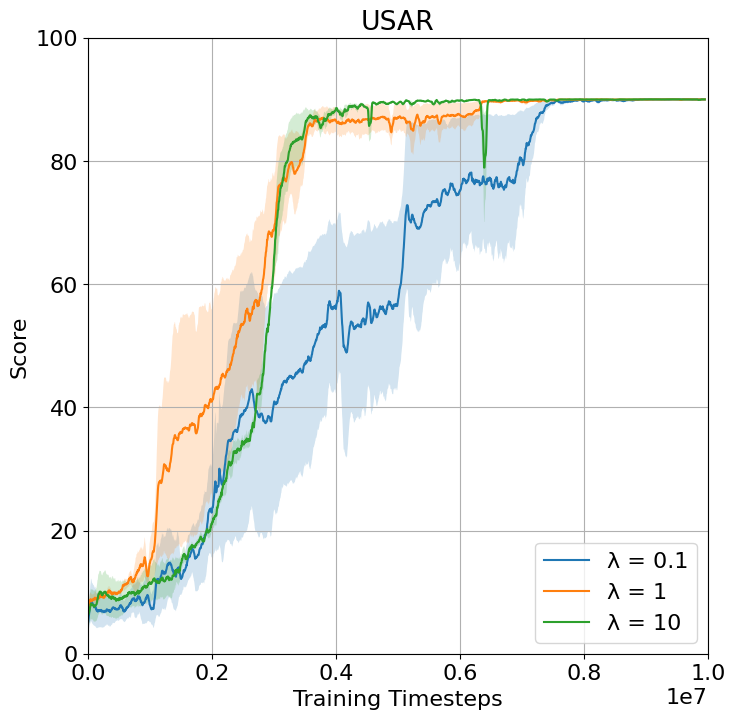}
    \captionof{figure}{Impact of supervised learning loss weight $\lambda$ on team performance.}
    \label{weight_perf}
  \end{minipage}
    \hfill
  \begin{minipage}[b]{0.32\textwidth}
    \centering
\includegraphics[width=\textwidth]{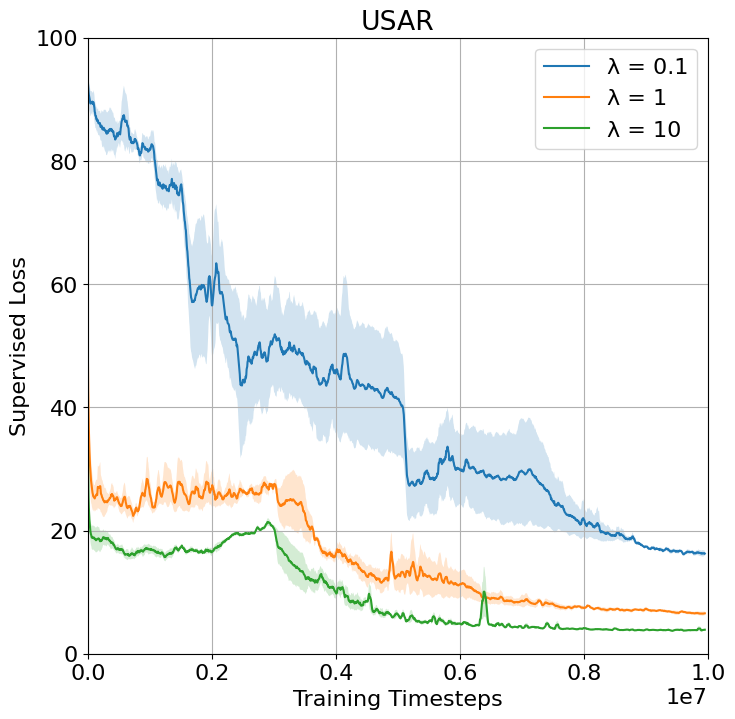}
    \captionof{figure}{Impact of SL loss weight $\lambda$ on communication alignment.}
    \label{weight_loss}
  \end{minipage}
\end{figure}

The MARL-comm agent uses a gating function to learn whether to communicate at specific timestamps. We could ablate this function to see its impact on team performance. As shown in the Fig.~\ref{removeGating}, removing the gating function harms the performance of LangGround more than IC3Net. This means both RL and SL signals influence the content and timing of LangGround communication.

In addition, we could manipulate the weight of supervised learning loss, i.e. $\lambda$ in function~\ref{weightedLoss}, to illustrate the contribution of RL and SL signals. As shown in Fig.~\ref{weight_perf} and ~\ref{weight_loss} in the PDF, $\lambda$ matters for both task performance and supervised loss. If the SL loss is weighted too high, the LangGround agent cannot optimize its policy in completing the task. While if the RL loss is weighted too high, the LangGround agent cannot align its communication with human language. This result aligns with our claim that RL optimizes the communication for task utility and SL optimizes the communication for alignment.

\end{document}